\documentclass[amsmath,amssymb,amsfonts,aps,prstper,citeautoscript,10pt,twocolumn]{revtex4-1}
\usepackage{graphicx}
\usepackage{multirow}
\usepackage{booktabs}

\begin{document}
\title{Integrated Testlets and the Immediate Feedback Assessment Technique}
\author{Aaron D. Slepkov}
\email{aaronslepkov@trentu.ca}
\affiliation{Trent University, Department of Physics and Astronomy, Peterborough, ON K9J 7B8, Canada}

\begin{abstract}
Multiple-choice testing offers attractive procedural advantages in formal classroom assessments, making this technique a popular tool in a wide range of disciplines. The use of multiple-choice (MC) questions in introductory-level physics final exams is relatively limited, largely due to reservations about its ability to test the broad cognitive domain that is routinely accessed with typical constructed-response (CR) questions. Thus, there is a need to explore ways in which MC questions can be utilized pedagogically more like CR questions while maintaining their attendant procedural advantages. In this article we describe how an answer-until-correct MC response format allows for the construction of fully multiple-choice examinations designed to operate much as a hybrid between standard MC and CR testing. With this tool---the immediate feedback assessment technique (IF-AT)---students gain complete knowledge of the correct answer for each question during the examination, and can use such information for solving subsequent test items. This feature allows for the creation of a new type of context-dependent item sets; the ``integrated testlet''. In an integrated testlet certain items are purposefully inter-dependent and are thus presented in a particular order. Such integrated testlets represent a proxy of typical CR questions, but with a straightforward and uniform marking scheme that also allows for granting partial credit for proximal knowledge. As proof-of-principle, we present a case study of an IF-AT-scored midterm and final examination for an introductory physics course, and discuss specific testlets possessing varying degrees of integration. In total, the polychotomously-scored items are found to allow for excellent discrimination, with a mean item-total correlation measure for the combined 45 items of the two examinations of $\overline{r}'=0.41\pm 0.13$ (mean $\pm$ standard deviation) and a final examination test reliability of $\alpha=0.82$ ($n=25$ items). Furthermore, partial credit is shown to be allocated in a discriminating and valid manner in these examinations. As has been found in other disciplines, the reaction of undergraduate physics students to the IF-AT is highly positive, further motivating its expanded use in formal classroom assessments. 

\end{abstract}

\maketitle
\section{Introduction}\label{sec:intro}
A typical final exam in an introductory physics course consists of a mixture of constructed-response (CR) exercises and problems. More recently, the addition of a multiple-choice (MC) question component is becoming increasingly common \cite{Aubrecht,Scott,Tobias}. The main reasons for this include limitations on instructor grading time, and limited financial resources for paid grading. The procedural advantages of using MC questions over CR questions includes simplified scoring that is both more reliable and considerably less labour intensive \cite{Scott,Haladyna2004}. It is, however, being recognized that with proper construction MC questions are powerful tools for the instruction and assessment of conceptual physics knowledge \cite{Aubrecht,Hestenes},  and there are examples of introductory physics final exams that consist entirely of MC questions \cite{Scott}. These, however, tend to be in universities with large class sizes, where the procedural advantages of MC questions are weighed against any pedagogical disadvantages stemming from an examination that largely measures compartmentalized conceptual knowledge and rudimentary calculation procedures. MC questions are not typically used to assess the complex combination of cognitive processes needed for solving numerical problems that integrate several concepts and procedures. Those kinds of problems involve the integration of a sequential flow of ideas---a physical and mathematical argument of sorts---that seems to resist compartmentalization.  For these reasons MC questions usually make up a relatively small portion of formal assessments, where greater weight is placed on traditional CR questions that involve problem solving and explicit synthesis. Thus, in order to broaden the utility of MC testing as a complete assessment tool in introductory physics education, we need to explore ways of using it more like CR testing while keeping the attendant advantages of both assessment types. In this paper we present a practical strategy for using MC questions to assess students' abilities to solve the kinds of complex numerical problems that are typically confined within CR formats.

Context-dependent item sets \cite{Haladyna1992}---or ``testlets"\cite{Wainer,Sireci}---can be utilized as key tools for creating a flow of ideas in a multiple-choice test. A traditional testlet comprises a group (two or more) of context-dependent MC items that are developed together to test a particular topic or area of knowledge. Testlets are used for a variety of reasons across disciplines: For example, because items within a testlet
\cite{Haladyna1992} share a common stimulus, their use reduces the amount of required reading and processing as compared to an equivalent number of stand-alone questions. This, in turn, allows for more questions to be used in a fixed-time exam and thus helps to improve test reliability and knowledge coverage. A reading comprehension testlet provides a classic example, in which a passage is provided and then several subsequent MC questions are used to probe the student's comprehension of ideas within the passage \cite{Sireci}.  In physics examinations, testlets often consists of a single diagram and the description of a physical scenario, followed by a group of questions that \emph{independently} probe the understanding or procedural knowledge pertaining to different concepts tied to that scenario \cite{Scott}. Testlets and CR questions are similar in that they both share a common scenario which is to be subsequently analysed in a multifaceted manner. Thus, testlets may serve as a proxy for traditional CR questions. However, traditional testlets differ from CR questions in an important way: Whereas solving CR physics problems requires the \emph{integration} of key concepts and procedures, MC testlet items are \emph{designed to be independent} of each other. In fact, item independence---the opposite of integration---has been a key attribute of testlet design theory \cite{Haladyna1992}. One main reason why MC testlets are constructed with independent items is the need for fair test scoring. In integrated sets of questions, the solutions may build upon each other in a sequential manner. Because MC questions are typically scored dichotomously (full credit for correct choice; zero otherwise), and because the test-taker has no knowledge of their results at each step, it is impractical and unfair to score an inter-dependent set of questions in a traditional MC test. On the other hand, in CR answers the scorer often has far more contextual information in the generated solution and thus is better able to glean the student's thought process and assign partial credit for work that is technically incorrect but conceptually or procedurally sound.

Here, development of ``integrated testlets''---in which some items are inter-dependent sequentially---may provide a bridge between traditional MC and CR questions. In an integrated testlet, one task may lead to another procedurally, and thus the knowledge of how various concepts are related can be assessed. This approach represents a markedly different way of using testlets. For example, whereas the items in traditional testlets (see for example questions 21-24 in Scott et al. \cite{Scott}) can be presented in any order, the items in an integrated testlet are best presented in a particular sequence. This kind of approach, however, could be grossly unfair and unpopular because student error in the first item would necessarily propagate through many other items, leading to multiple jeopardy. The most directly viable means of using integrated testlets may thus requires the test-taker to have immediate confirmatory or corrective feedback for each item, thereby allowing the participant to gauge---and if necessary to modify---their approach before each step \cite{Baldwin}. Likewise, such immediate feedback would allow for fair exam scoring whereby each concept or procedural step is assessed independently.

The requirements for immediate feedback can be easily satisfied by the use of computers-administered exams. However, despite two decades of widely-accessible computers, most university examinations are still conducted in traditional classroom settings \cite{Peat}.  A relatively new type of in-classroom MC response format known as the Immediate Feedback Assessment Technique, or IF-AT, \cite{Epstein2002-52,DiBattista2005} has been designed to allow for confirmatory and corrective feedback. The IF-AT is a commercially-available ``scratch-and-reveal''-type MC answer form. The IF-AT response sheet consists of rows of bounded boxes, each covered with an opaque waxy coating similar to those on scratch-off lottery tickets (see Fig. 1). Each row represents the options from one MC question. For each question, there is only one keyed answer, represented by a small black star under the corresponding option box. Students make their response by scratching the coating off the box that represents their chosen option. If a black star appears inside the box, the student receives confirmation that the option chosen is correct, and proceeds to the next question. On the other hand, if no star appears, the student immediately knows that their chosen option is incorrect. The student can then reconsider the question and continues scratching boxes until the star indicating the keyed option is revealed. Thus, the IF-AT is also known as an answer-until-correct assessment technique.

The IF-AT possesses several properties that make it an attractive pedagogical tool: For example, the immediacy of the feedback provided by IF-AT has been shown to promote learning over other assessment techniques that can at best only provide delayed feedback \cite{Dihoff}. Because the students learn what the correct answer is, rather than just learning their score on a given question, an IF-AT exam then becomes a learning opportunity within the auspices of an assessment \cite{DiBattista2005}. Furthermore, this answer-until-correct method allows for a variety of scoring schemes for a response sequence where a student initially provides an incorrect response, but then responds correctly after reworking the problem \cite{DiBattista2009}. This, combined with the full knowledge of results, leads to a greater sense of fairness in students, who nearly universally prefer this technique to other forms of MC assessment \cite{DiBattista2004}. The ability to either confirm or correct student responses in real time (without additional instructor resources) makes IF-AT an enabling tool for the development and administration of integrated testlets. The combined application of these two tools has the potential to significantly expand the way MC testing is implemented in the physical sciences and beyond.

In this proof-of principle article we address a simple set of preliminary research questions: Can the IF-AT be used to develop testlets with varying degrees of integration that can be effectively administered in a formal physics classroom setting? Furthermore, is the partial-credit provided by the simple and automatic scoring scheme allotted in a fair and discriminating manner? Finally, what are some key structural and procedural considerations to creating integrated testlets with the IF-AT? In answer to these questions, we describe the preparation and functionality of a pair of MC exams deployed in a calculus-based introductory electricity and magnetism course. These exams comprised both stand-alone MC questions, and MC questions within testlets with varying degrees of item integration. We describe the administration of these tests using the IF-AT, which allows students to obtain item-by item confirmatory or corrective feedback, and which allows for scoring schemes that incorporate partial credit. We then analyse the quality of the MC items on the exams, the reliability of the exams, and the validity of granting partial credit. We find that, indeed, partial credit is found to be granted in a discriminating fashion. Finally, using a set of testlet examples, we discuss important considerations relevant to both the construction of integrated testlets and to the successful use of the IF-AT in physics education.

\begin{figure*}
 \includegraphics[width=0.75\textwidth]{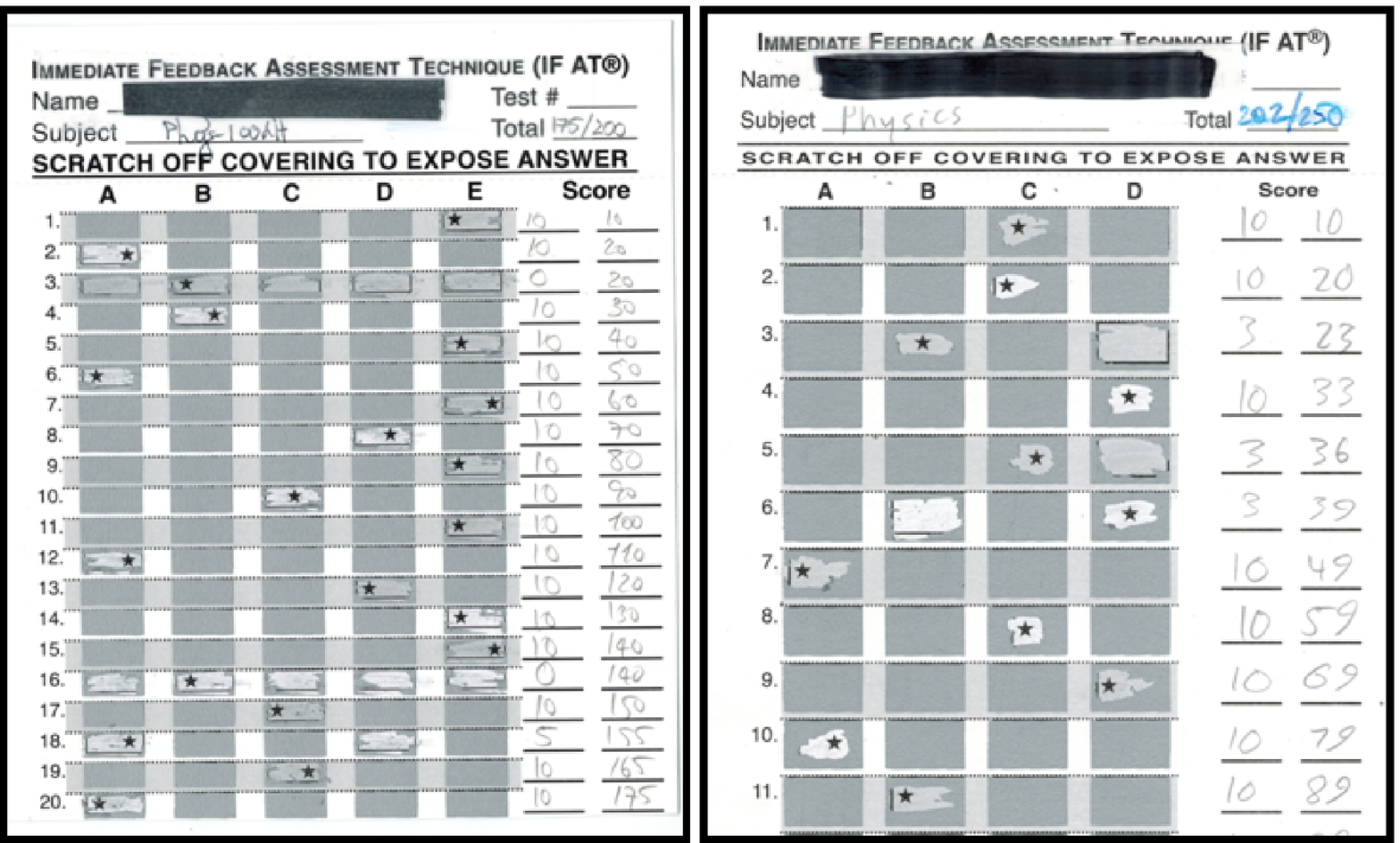}
 \caption{Examples of used IF-AT examination forms. (Left panel) A two-hour midterm consisting of twenty 5-option questions was given in which full credit (10) was given for answers correct on the first try, partial credit (5) was given for answers correct on the second try, and no credit for answers correct on subsequent tries. This is denoted a [10,5,0,0,0] marking scheme. (Right panel) A three-hour final consisting of twenty-five 4-option questions was given where the marking scheme was [10,3,0,0]. Only part of the 25-item form is shown. The students may grade the forms themselves as they proceed along the test.}\label{fig:IF-ATcards}
\end{figure*}

\section{Methods}
\subsection{Course design}
A one-term calculus-based course titled ``Introductory Physics II---Electricity and Magnetism'' was offered in 2012 at a primarily undergraduate Canadian university. 60 students were enrolled at the start of the course. The course is a requirement for physics and chemistry majors, and covers topics such as electro- and magneto-statics, simple circuits, introductory quantum physics, and optics. Course delivery followed peer-instruction and interactive-learning principles \cite{Mazur,Crouch,Meltzer}: Both a textbook and departmentally-written course notes were adopted and students were assigned pre-class readings from these sources. A computer-graded just-in-time (JIT) online reading quiz was administered before each class, at the student's leisure, wherein the instructor also solicited information regarding misunderstandings and difficulties in the assigned readings \cite{Novak}. In-class instruction consisted of 5-10 minute miniature expositions and summaries arising from JIT questions, followed by clicker-based conceptual tests and group discussion. Bi-weekly laboratory sessions were alternated with bi-weekly recitation sessions at which knowledge of assigned problem-set answers was tested with 45-minute constructed-response quizzes, followed by tutoring of the next problem-set.

A two-hour midterm examination comprising 20 MC questions was administered on week 8 of the 12 week term, with student responses recorded using the IF-AT.  A three-hour final examination consisting entirely of 25 MC questions was also administered using the IF-AT response forms. A detailed formula sheet was provided to the students at both examinations. Examinations were collectively worth 50\% of a student's final grade. In total, 51 students wrote the mid-term and 49 students wrote the final exam. 

\subsection{Construction of exams}
Typical multiple-choice-based final examinations in introductory courses (such as biology, chemistry, or psychology, for example) use a large number of MC items in order to cover a wide breadth of course material. This often means constructing exams with over one hundred questions \cite{DiBattista2011}. For the exams considered here, the MC format was used for a range of conceptual, analytical, and calculation-based questions, each of which may require more attention and time than a traditional MC question. Thus, only 20 questions were used for a two-hour examination or 25 questions for a 3-hour examination. This is half as many questions per hour as typically used by others in similar courses \cite{Scott}. To assure adequate and efficient coverage of course material, an examination blueprint was prepared, as is recommended for valid assessment construction \cite{Aubrecht}. The writing of the individual questions also followed recommended MC item construction principles and practices that are meant to maximize item discrimination and test reliability \cite{Aubrecht,Haladyna2004}. It should be noted that with the IF-AT, the answer key is immutably built into the scratch-cards and thus the MC questions need to be constructed to match \cite{MultipleKeys}. This also means that the IF-AT is less forgiving of minor errors than other MC techniques. Thus, to aid the proper construction of the tests, the mid-term and final examinations were ``test-driven'' by teaching assistants before being administered to the class.

In our course, the IF-AT technique was adopted primarily as a tool for exploring the viability of integrated-testlets in formal course assessments. Thus, the midterm and final exam were constructed to have a balance between such testlets and stand-alone MC questions. The 20-question midterm consisted of 4 stand-alone MC questions and 4 testlets that together comprised 16 MC questions. The 25-question final examination contained 8 stand-alone MC questions and 5 testlets comprising 17 MC questions.

The exam scoring was designed for simplicity, and all items were worth an equivalent number of marks. The midterm examination used twenty 5-option items, wherein full credit (10) was given for an initial correct response, and half-credit (5) was given for items correct on the second attempt. No credit was earned for subsequent attempts. For such a test the marking scheme can be defined as [1.0,0.5,0,0,0] and the expected mean test score from purely guessing students is 30\% (calculated as $0.2\times 1 + 0.8\times 0.25 \times 0.5 = 30\%$). For such a 20-question test there is only a 4.8\% chance that a student can pass the test (i.e. obtain $>$50\%) by guessing alone. The final exam used twenty-five 4-option items, wherein credit (10) was given for an initial correct response, and partial credit (3) was given for items correct on the second attempt. For this [1.0,0.3,0,0] test, the expected mean test score from purely guessing students is 32.5\%, and there is only a 3.4\% chance that a student can pass the test by guessing alone. The partial credit scoring schemes were chosen as a compromise between the likelihood of passing the exam by chance and the anticipated effect of sustained incentives. That is, a chance score of 30\%-35\% was balanced by the perception that students will be suitably motivated to think (rather than guess) on a second try, but perhaps not so on a third try. Others who utilize the IF-AT routinely use scoring schemes that give partial credit beyond the second try \cite{DiBattista2009}. In fact, as a result of this study, we have effectively used a [1,0.5,0.1,0,0] scoring scheme in several subsequent exams.  It should be noted that with the IF-AT students can tally their own exam score on the answer form, and most students take the opportunity to do so (see Fig.~\ref{fig:IF-ATcards}).

\section{Results and discussion}
Midterm and final exam scores were 67\% $\pm$ 18\% (mean $\pm$ standard deviation) and 75\% $\pm$ 14\%,  respectively. The scores ranged from a minimum of 45\% to a maximum of 98\% on the midterm and from 28\% to 100\% on the final examination. The overall class grade for students who completed the course was 70\% $\pm$ 16\%, with scores ranging from 37\% to 98\%.

A statistical item analysis of the MC questions helps to put their functionality in context with other MC tests. Traditional test analysis theory looks at three main aspects of the test; individual item analysis with respect to item difficulty and discrimination, test reliability, and test validity \cite{Ding_PRSTPER_2009,Haladyna2004,Ebel}. A summary of test analysis metrics is provided in Table 1.

\subsection{Difficulty and discrimination}
The more difficult an item, the lower the proportion of available marks that will be earned by the students. A widely-used item difficulty parameter, $p$, is defined as the mean obtained item score. Typically in MC test analysis the scoring is dichotomized and $p$ is simply the proportion of the students that answer the question correctly. In our use of the IF-AT, where partial credit is allocated for being correct on subsequent selections, a continuous or polychotomous difficulty parameter $p'$ can be defined to represent the obtained mean item score. Both $p$ (and $p'$) range between 0 and 1, and the value of each decrease with item difficulty. For example, for a [1.0,0.3,0,0] test in which 1/3 of the class answer correctly on the first selection, and 1/3 of the class answer correctly on the second selection, the values of $p$ and $p'$ would be 0.33 and 0.43, respectively.

The various items on the two exams ranged widely in difficulty, with the easiest and most difficult questions measuring $p'=0.92$ and 0.32, respectively, as presented in Table~\ref{tab:Metrics}.

\begin{table*}[ht]
\caption{Test item analysis and summary metrics for the midterm and final 				examinations}
\bgroup
\def\arraystretch{1.5}
{\setlength{\tabcolsep}{0.5em}
\begin{tabular}{  l | c || c | c  }
 \hline \hline
\textbf{Test metric} & \textbf{Parameter} & \textbf{Midterm Exam} & \textbf{Final Exam} \\
 \hline\hline
 \textbf{\# items} & \emph{n} & 20 & 25 \\
 \hline
\textbf{ \# students} & \emph{N} & 51 & 49 \\
 \hline
 \textbf{item difficulty}\footnote{Item difficulty. This is the mean of the item achievement scores. $p$ is the dichotomized parameter and $p'$ includes partial credit. M = mean; SD = standard deviation; min = minimum value; max = maximum value.} & $\overline{p}$ & 0.67 $\pm$ 0.17; 0.29\ldots 0.86 & 0.62 $\pm$ 0.15; 0.27\ldots 0.88 \\
 \cline{2-4}
 M $\pm$ SD; min\ldots max & $\overline{p}'$ & 0.75 $\pm$ 0.16; 0.36\ldots 0.92 & 0.67 $\pm$ 0.14; 0.32\ldots 0.91 \\
 \hline
\textbf{item-total correlation}\footnote{A measure of item discriminatory power, the item-total correlation is the point-biserial correlation in the case of dichotomous scoring ($\overline{r}_{i-t}$) and the Pearson-$r$ when partial credit are included ($\overline{r}_{i-t}'$). M = mean; SD = standard deviation; min = minimum value; max = maximum value.} & $\overline{r}_{i-t}$ & 0.36 $\pm$ 0.13; 0.12\ldots 0.55 & 0.42 $\pm$ 0.14; 0.13\ldots 0.68 \\
 \cline{2-4}
M $\pm$ SD; min\ldots max  & $\overline{r}_{i-t}'$ & 0.39 $\pm$ 0.13; 0.15\ldots 0.60 & 0.43 $\pm$ 0.13; 0.16\ldots 0.69 \\
\hline
\multirow{2}{*}{\textbf{test reliability}\footnote{Test reliability measure. In the non-dichotomized set, this is Cronbach's alpha ($\alpha$). $\alpha_{50}$ is the value of $\alpha$ when adjusted for comparisons with a 50-item test.}} & $\alpha$ & 0.71 & 0.82 \\
\cline{2-4}
& $\alpha_{50}$ & 0.86 & 0.90 \\
\hline
\end{tabular}}
\egroup
\label{tab:Metrics}
\end{table*}

The spread of item difficulties is not a problem in of itself, and was in fact obtained largely by design in an attempt to utilize less-difficult questions as occasional confidence boosters, especially early in an exam or mid-testlet. Considering both examinations, the intra-testlet items were slightly more difficult than the stand-alone questions, with $p'=0.69\pm0.15$ and 0.75 $\pm$ 0.15, respectively.

More important than item difficulty is the power of a given question to discriminate between more knowledgeable and less knowledgeable students. Whether an item is relatively easy or difficult may be immaterial as long as the item is properly discriminating. Several parameters are commonly used for measuring the discriminatory power of test items, including the extreme-groups \emph{item-discrimination index}, and the item-total point-biserial correlation or \emph{item-discrimination coefficient} \cite{Ding_PRSTPER_2009,Haladyna2004,Ebel}. The point-biserial (PBS) correlation is simply the Pearson-$r$ correlation measure for data in which one parameter is dichotomous. When analysing our exam results, we can consider either only first-responses, thereby dichotomizing the data and better approximating how the exams might operate from a standard MC format perspective, or we can include the partial-credit allotted by the answer-until-correct format. We denote the item-total correlation parameters that measure the discriminatory power of the test items as $\overline{r}_{i-t}$ and $\overline{r}_{i-t}'$, respectively depending on whether they include only first-response (dichotomous scoring) or all responses (polychotomous scoring). Conventional wisdom holds that items with a discrimination coefficient below 0.2 are insufficiently discriminating and should be either modified or discarded in subsequent test iterations \cite{Haladyna2004}. Dichotomizing our data, we find an excellent mean item-discrimination coefficient of $\overline{r}_{i-t}= 0.39\pm 0.14$ across both examinations, with individual items ranging in value from 0.12 to 0.68 (see Table~\ref{tab:Metrics}).  For polychotomous scoring (as was actually used in the course) we obtain a mean value of $\overline{r}_{i-t}'= 0.41\pm 0.13$ with individual items ranging in value from 0.15 to 0.69. The fact that $\overline{r}_{i-t}'$ is greater than $\overline{r}_{i-t}$ supports the notion that partial credit was used effectively in the IF-AT exams (see validity discussion, below). Only 4 out of 45 items had a polychotomous discrimination coefficient smaller than 0.2. Overall, the average level of discrimination we obtained is exceptional for a set of classroom tests, \cite{DiBattista2011} and shows that, at the very least, the IF-AT can be used to properly administer discriminating item sets with a wide range of difficulty levels, both in stand-alone items and in integrated-testlets. 

\subsection{Reliability}
It is possible for a test to measure many things in parallel. An examination in an introductory physics class should strictly assess the students' knowledge of the subject matter. It is difficult to construct/deconstruct standard physics problems in such a way that they only test physics knowledge; there is always some collateral testing of extra-disciplinary skills such as mathematical ability, reasoning, time management, etc. Demonstration of the integration of these skills within a broader framework of physics knowledge is actually quite desirable in a typical physics exam, but it must be done in a consistent and uniform manner. This is at the heart of what is meant by reliability. A common measure of internal consistency is the Cronbach's alpha ($\alpha$) which is applicable for both dichotomous and non-dichotomous data \cite{Ding_PRSTPER_2009,Haladyna2004,Ebel}. The value of $\alpha$ can range between 0 (for an utterly unreliable test) to 1.0 (for a perfectly reliable test). Traditionally, a test yielding $\alpha<0.7$ is considered relatively unreliable, while a value above 0.8 is considered very good and a value above 0.9 is considered excellent. The midterm and final examinations in this course yielded $\alpha=0.71$ and 0.82, respectively. Generally, reliability of a test is expected to improve with an increasing number of items. Thus, in order to compare the reliability between tests of different lengths, it is common to adjust the reliability coefficient to reflect that of a standard test length; typically of 50 questions \cite{Aubrecht,Bonder}. When our exam reliabilities are adjusted to correspond to a test length of 50 items (see Reference  \cite{Bonder} for formulas), our obtained values of $\alpha_{50}=0.86$ and 0.90 allow for comparisons with a wide group of other MC tests, \cite{DiBattista2011} and are found to be very reliable for classroom exams. 

\subsection{Validity}
\begin{figure}
 \includegraphics[width=\columnwidth]{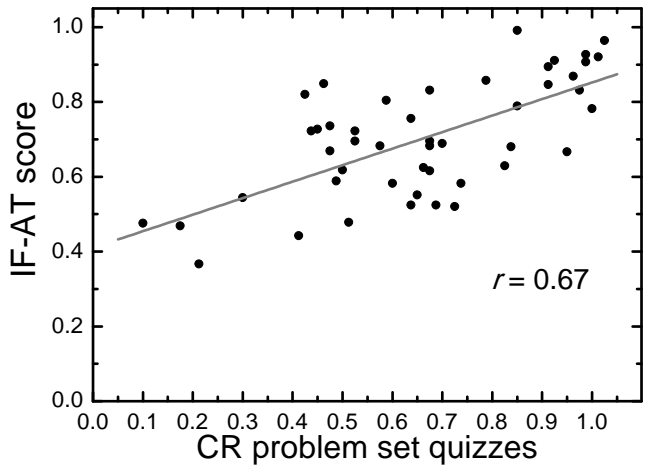}
 \caption{Correlation between total score on the IF-AT items from final and midterm exams and score on bi-weekly constructed-response quizzes. The straight line is from a linear regression, obtaining a correlation with value $r=0.67$. A positive correlation between achievement in constructed-response and IF-AT questions bolsters the conclusion that the IF-AT exams are \emph{valid} assessments of learning outcomes of the course.}\label{fig:CR-IF-ATcorrelation}
\end{figure}

The concept of validity is an evaluation of how appropriately the content and functionality of the test measures the trait that is to be assessed \cite{Ebel,Allen}. In the case of the examinations under consideration here, it is an estimation of how well these tests assess students' knowledge of the specific course material as a whole. To assure content validity, the final examination covered the broad range of course content, and was constructed with the aid of a blueprint that helped avoid putting too much weight on a limited set of concepts and cognitive tasks \cite{Aubrecht}. This blueprint was used for holistic alignment of the exam to course content, and while it guided testlet selection it did not guide individual testlet construction. Establishing content validity of classroom examinations can prove difficult, as it is closely tied to the instructional and assessment histories in the course. Thus, an appropriate indication of overall validity may be gauged by a correlation between students' cumulative achievement on the IF-AT-administered examinations and their achievement in related course material. A good example of such material may be found in the scores of five bi-weekly quizzes administered in recitation sessions. Each of these quizzes comprised two constructed response questions selected by the recitation leader from the assigned problem sets. The quizzes were graded by the recitation leader, whom had only a limited role in the creation of the IF-AT examinations. Figure~\ref{fig:CR-IF-ATcorrelation}
displays a scatter plot of student examination scores against student achievement in the bi-weekly constructed-response quizzes. With 47 participants who both completed course materials and wrote the final exam, the correlation ($r=0.67$) is sufficiently clear to establish the validity of this particular exam as a proper assessment of student knowledge of course material, and is well in line with similar comparisons for validated MC tests \cite{Pfeiffenberger_1977}. In previous course implementations with similar course delivery structure but traditional CR examinations, this correlation between the bi-weekly quizzes and final examination score ranged from $r=0.35$  to $0.73$, with no concerns having been raised regarding the validity of the quizzes or examinations in those courses. Furthermore, it should be mentioned that in the behavioral sciences, a correlation of $r > 0.5$ represents a
``strong effect" \cite{Cohen}. 

\subsection{Allocating partial credit in a multiple-choice test}
The issue of partial credit is important within the context of classroom assessments in physics. Students nearly universally feel that allocation of partial credit is essential for a fair marking scheme; particularly for computational tasks. However, some instructors are loath to assign partial-credit for answers that are ultimately incorrect, despite the suspicion that students have partial knowledge. This has been famously referred to as the dead-mouse problem, wherein a solution that uses parts-wise correct elements may still be utterly unacceptable in context of how those elements are put together \cite{Rogers}. Nonetheless, many scoring schemes in constructed-response tests include a quasi-continuous or holistic scale of partial credit \cite{Spurgin,Johnson}. There are even proposed MC scoring schemes that provide partial credit based a pre-determined value scale of how correct (or, ultimately incorrect) various responses are \cite{Lin_WeightedMCscoring_2012}.

Finding a valid way to assess partial knowledge in MC testing has proven problematic for many reasons, including the fact that scantron\textsuperscript{\textregistered}-type answer forms are designed around single-choice selection and dichotomous scoring. Nonetheless, several scoring schemes have been developed to assess partial knowledge in MC tests \cite{Chang,Ben-Simon},  including elimination testing \cite{Chang}, probability scoring \cite{Ben-Simon}, subset selection testing, \cite{Chang,Jaradat}, distractor weighting \cite{Lin_WeightedMCscoring_2012}, and distractor ordering \cite{Ben-Simon}. While some of these schemes have been demonstrated to improve test reliability over the traditional dichotomous ``number-correct'' marking scheme \cite{Ben-Simon,Rodrigues}, most are relatively convoluted and add considerable complexity to either the scoring or test taking. A major advantage of the IF-AT is that its answer-until-correct format allows for simple hands-off integration of partial credit marking schemes \cite{DiBattista2009}. The most commonly used marking scheme with IF-AT involves granting of full credit for questions answered correctly on the first attempt, followed by diminishing (positive) partial credit for correctly answering on subsequent trials. For our examinations, students only received partial credit for answers correct on the second attempt, with no credit being given for correct responses on the third and fourth attempts. This scheme is easy to implement both for the student and for the grader. 

There are some key differences in how partial credit is granted within the IF-AT, as compared to other question types, such as CR. One may consider that while partial credit in IF-AT is given for proximal knowledge \cite{Epstein2001}, it is actually being given for the route in which the student arrives at the correct answer. However, in what may be a subtle (but hopefully not esoteric) argument, the student always gets the credit for coming up with the completely correct answer and not for a partially correct answer. The final response of the student is always the fully correct one; it is not left to the instructor to interpret what elements of the student's answer are sufficiently correct to merit partial credit. In this marking scheme an implicit valuation is being made that once a student makes a mistake, that this mistake can be corrected, learning can take place, and a fully correct answer can then be made anew. This marking scheme takes into account the various opportunities granted to the students to display their knowledge, but ultimately, to get any credit the student needs to select the correct response and not a sub-optimal (dead-mouse) response. This point lies at the heart of how the IF-AT can be used as a tool that advantageously shares formative and achievement assessment properties \cite{Bennett2011}.

The fact that the IF-AT allows for incorporation of partial-credit does not, however, guarantee that the partial credit is being utilized in a valid or discriminating manner. An analysis of the IF-AT final exam scores strongly suggests that partial-credit was used effectively to determine knowledge. Overall, however, partial credit accounted for only a small proportion of the total exam score.  The mean score on the exam was 67\%, but without partial credit---i.e. a [1,0,0,0] marking scheme---the average was only 5 points lower at 62\%. Naturally, weaker students have a greater need for partial credit, and consequently partial credit makes up a larger proportion of their mark. In fact, as many as 10 of 49 students would have failed the final exam (earned less than 50\%) without partial credit, but partial credit accounted for less than 8 percentage points for each of these students. As expected, there is an inverse correlation between the amount of partial credit granted and the exam score. This is mostly due to opportunity; the top scorers are more likely to get full credit on any question and thus have fewer opportunities to earn partial credit. Nonetheless, the granting of partial credit proves discriminating. To demonstrate this, we consider the likelihood that a student earns \emph{available} partial credit. A student who answers an item correctly on their first response gets full credit and has no opportunity to earn partial credit. Only in cases when a first response is incorrect does a student have the opportunity to earn partial credit. When partial credit is used in a discriminating manner, we expect top students to earn a higher proportion of their available partial credit as compared to the students at the bottom. Indeed, the top fifteen final exam scorers \cite{WithoutTopTwo} earned 65\% $\pm$ 23\% (mean $\pm$ standard deviation) of the partial credit available to them, while the bottom 15 students earned only 39\% $\pm$ 10\% of the partial credit available to them. As confirmed by a t-test for independent variables, this is a statistically significant difference ($t\!\left(19.2\right)=-3.966;\, p \leq 0.001$). As the final exam consisted entirely of four-option MC questions, a purely-guessing cohort would be expected to garner 33\% of the available partial credit. Since students at all achievement levels on the (four-option) final examination are earning more than 1/3 of available partial credit, there does not appear to be a prevalence of pure guessing in second response attempts.  The ability to implement simple and valid partial-credit schemes in a MC test format is thus an important benefit of the IF-AT.

\subsection{Commentary on particular testlets and questions}
The multiple-choice examinations were designed to both cover a wide range of course topics and to assess student abilities across the broad cognitive domain. With the attendant benefits of immediate feedback tools one can create a variety of non-traditional MC questions and testlets as part of formal assessments in introductory physics classes. The following section demonstrates this fact and discusses the motivation and outcomes of testlets with varying levels of integration.

\begin{figure}
 \includegraphics[width=\columnwidth]{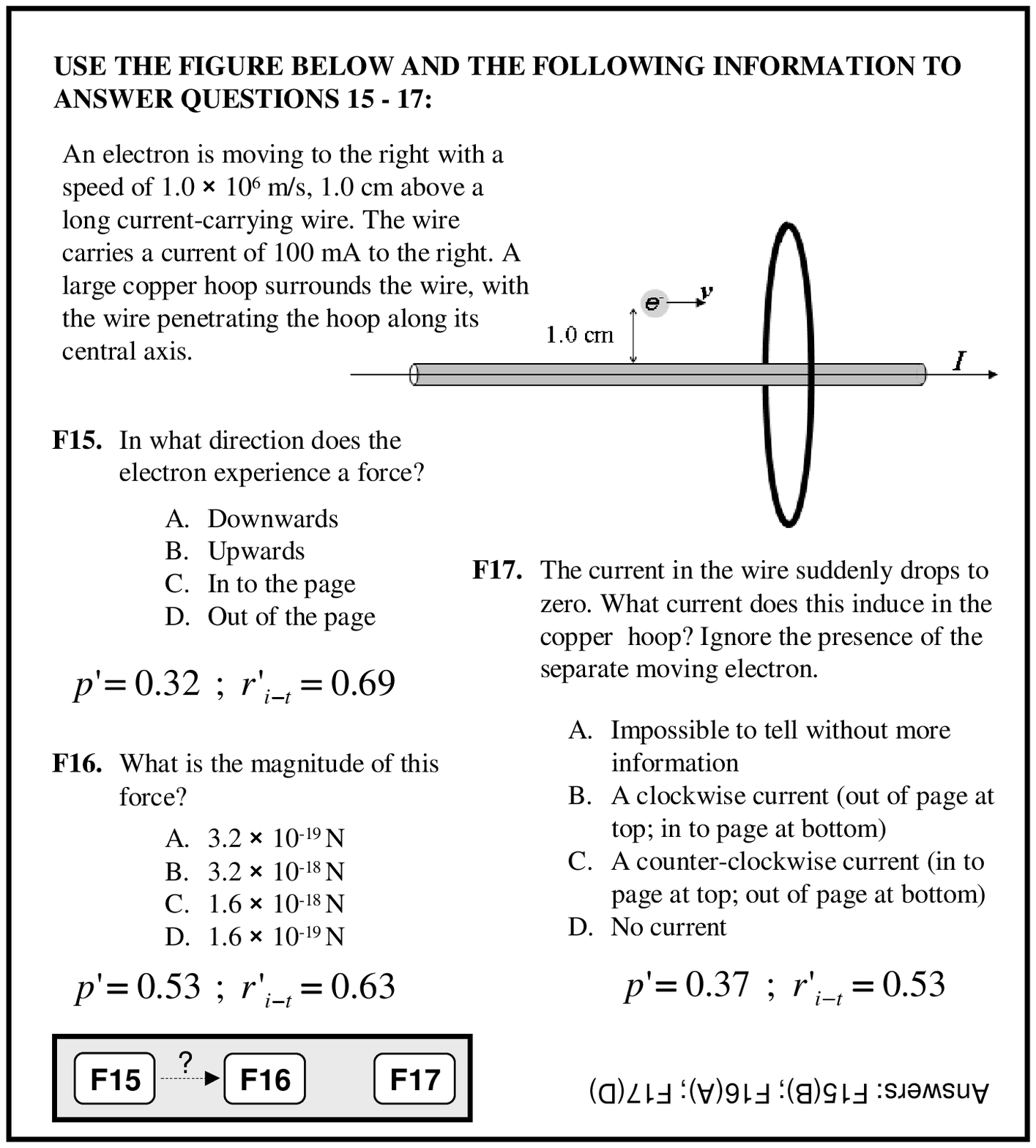}
 \caption{Three-item testlet comprising final exam questions F15-F17 as an example of a traditional testlet with minimal item integration. The dashed line with the question mark in the integration diagram at the bottom left indicates a possible (questionable) but very weak level of integration. Item F17, however, is completely separate from the others. The answers are provided upside-down in the bottom right. Neither the answers nor the integration diagrams were present on the actual examination papers. $p'$ is the item mean score with partial credit and $r_{i-t}'$ is the coefficient of correlation between item score and student exam score, representing a measure of item discriminatory power (See text for details).}\label{fig:F15-F17_NewSize}
\end{figure}

Traditional testlets are designed to contain as little integrated content as possible. Consider Figure~\ref{fig:F15-F17_NewSize}
which presents a three-item testlet comprising final examination questions F15, F16, and F17. This testlet is designed to have negligible levels of item integration, and thus, the order of the questions is expected to be relatively arbitrary \cite{QuestionOrder}. There is a weak link between questions F15 and F16, as the two are related, but the answer of one is not expected to impact the students' thought process in solving the other. Question F17 is completely separate conceptually from the other two. This testlet proves to be quite difficult, yet very discriminating. In fact F15 proves to be both the most difficult and most discriminating item on the final examination. This fact speaks volumes in support of the importance of testing conceptual understanding in an introductory physics exam. Its solution simply requires two consecutive implementations of ``the right hand rule'', combined with accounting for the sign of the charge in the Lorentz force vector.

\begin{figure}
 \includegraphics[width=\columnwidth]{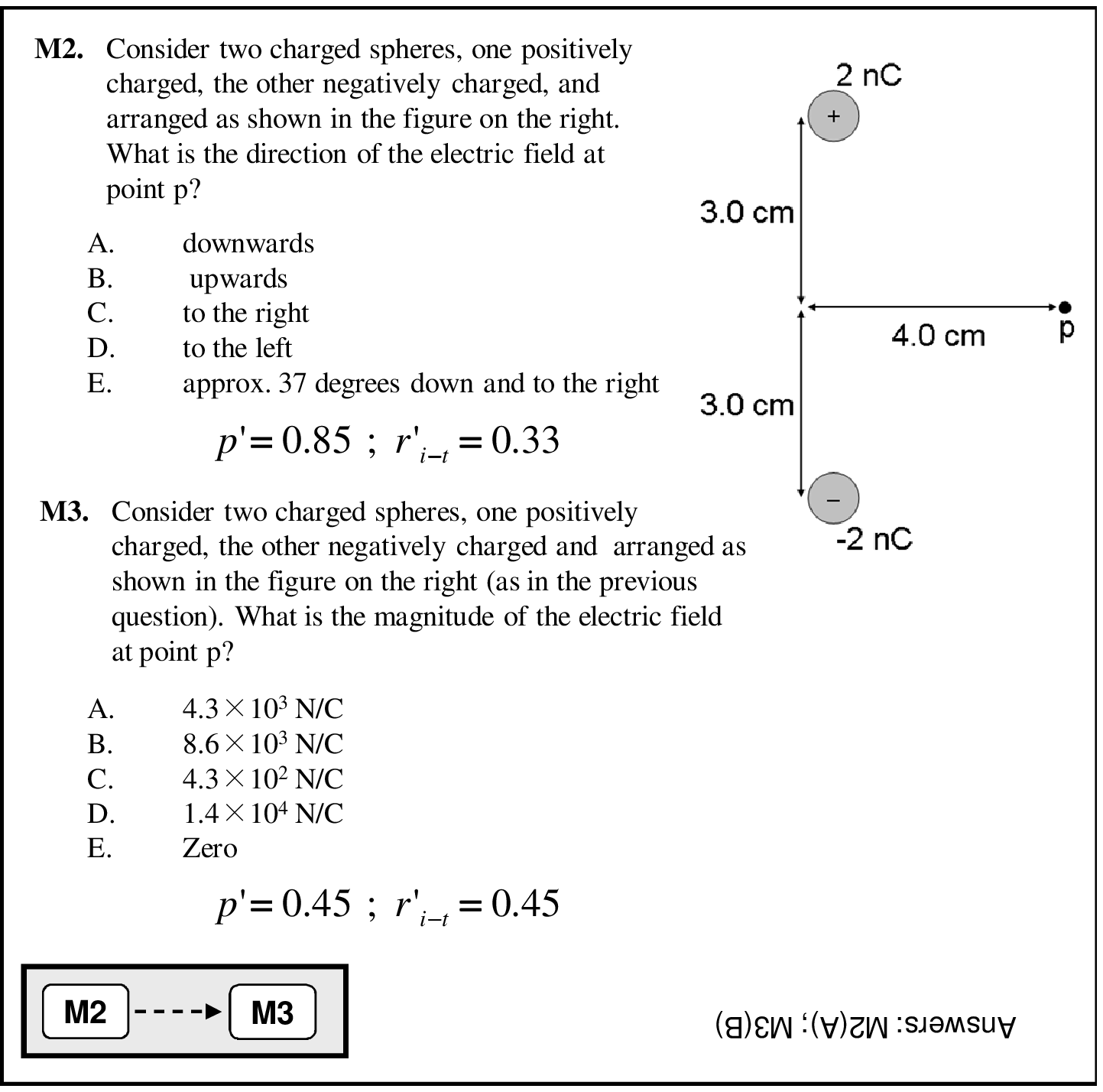}
 \caption{Two-item testlet comprising midterm questions M2 and M3 as an example of a partially-integrated item set. The dashed line in the integration diagram at the bottom left indicates a relatively weak integration. Here, the order of questions matters, and knowledge of the answer in M2 can provide insight and shortcuts for the calculation needed in M3.}\label{fig:M2-M3}
\end{figure}

Consider the two-item testlet presented in Figure~\ref{fig:M2-M3}
as an example of an item set with partial integration. Here, midterm questions M2 and M3 are only weakly interrelated. Full integration between two questions implies that knowledge of the answer of one is \emph{necessary} for answering the other.  However, when knowledge of the answer of one question simply aids the solution of the other, the questions can be considered partially integrated. In the case of M2 and M3, the solution to M2 is not strictly needed for a complete solution of M3, however, it is hoped that were M3 given as a CR question that students would first construct a solution to M2 as an intermediate step. Thus, knowledgeable students may thus find the order of questions in this testlet more helpful than would weaker students. In this testlet, it is expected that the order of questions matters, such that the order M2$\rightarrow$M3 makes answering M3 less difficult, but ordering the question M3$\rightarrow$M2 makes neither question less difficult.

\begin{figure}[t]
 \includegraphics[width=\columnwidth]{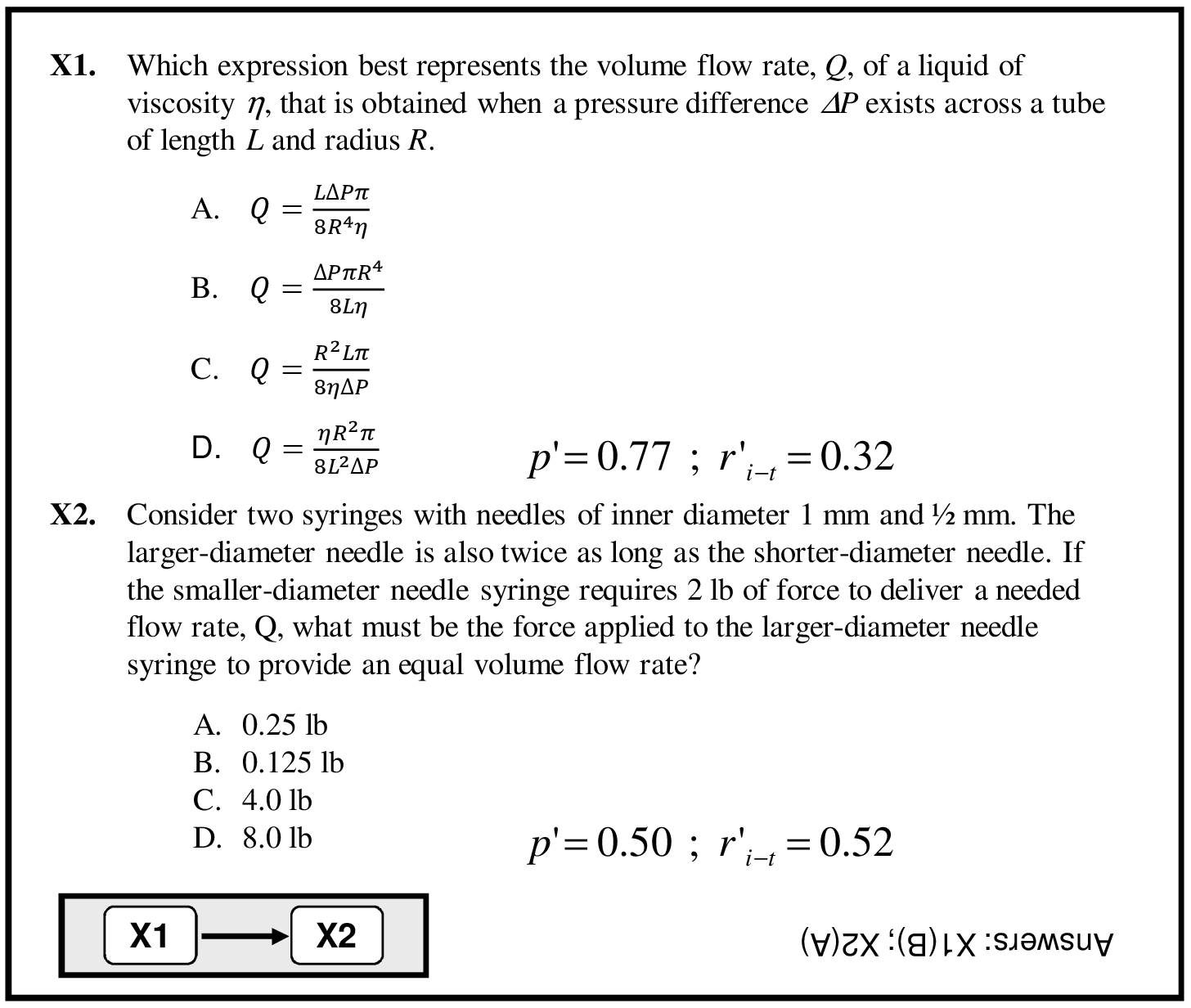}
 \caption{Two-item testlet comprising X1 and X2, taken from a [1.0,0.3,0,0] final examination  given to an Introductory Physics for the Life Sciences course. This testlet makes use of immediate feedback to ask an analytical question about an important formula (that has been redacted from the accompanying formula sheet). Once the students have reasoned the proper relationship between the variables in X1 (or used dimensional analysis), they are required to utilize the formula to answer X2. The questions are strongly integrated, as answering X2 \emph{requires} knowledge of the answer to X1.}\label{fig:X1-X2}
\end{figure}

Another good example of how the immediate feedback aspects of the IF-AT can be used to assess the type of knowledge we often find difficult to assess in a fair and sensible manner involves the analysis or derivation of important formulas. With the proliferation of formula sheets on final exams, it is often felt that students are becoming disconnected from the meaning of key equations. Rather, it is felt that students are simply being taught how to use those equations solely in a procedural manner. Instructors may wish to ask students to either derive, reason, or recall key relationships on the exam, but in doing so, the instructor is then unable to require the student to use that formula in subsequent questions for fear that success in the latter is too dependent on the former. This is a typical problem with any inter-related questioning on an exam. However, because students using the IF-AT can always obtain the correct answer for any question, there is little disadvantage in making such problems interrelated. Consider the two-item testlet shown in Figure~\ref{fig:X1-X2},
comprising two questions (X1 and X2) from a final exam given to an ``Introductory Physics for the Life Sciences'' course \cite{DiffMatSameMeth}. In X1 students are asked to identify Poiseuille's viscous flow-rate equation, and are then required to use that formula in a subsequent numerical question, X2. The corresponding formula was redacted on the formula sheet supplied with the examination papers. The identification of the correct relationship in X1 is not meant to be an exercise in recollection, but rather a question that allows students to combine various reasoning tools such as dimensional analysis and physical insight. This question proves to be relatively easy, yet it still has good discriminatory power. This item pair is stringently integrated, and must be presented in the provided sequence. Clearly, the order X2$\rightarrow$X1 would be nonsensical. Asking X2 without asking X1 is certainly viable, and would not be out of place as a CR question. However, the ability to include X1 provides the instructor a straightforward tool for assessing student understanding of how various parameters are interrelated in an important formula.

\begin{figure}
 \includegraphics[width=\columnwidth]{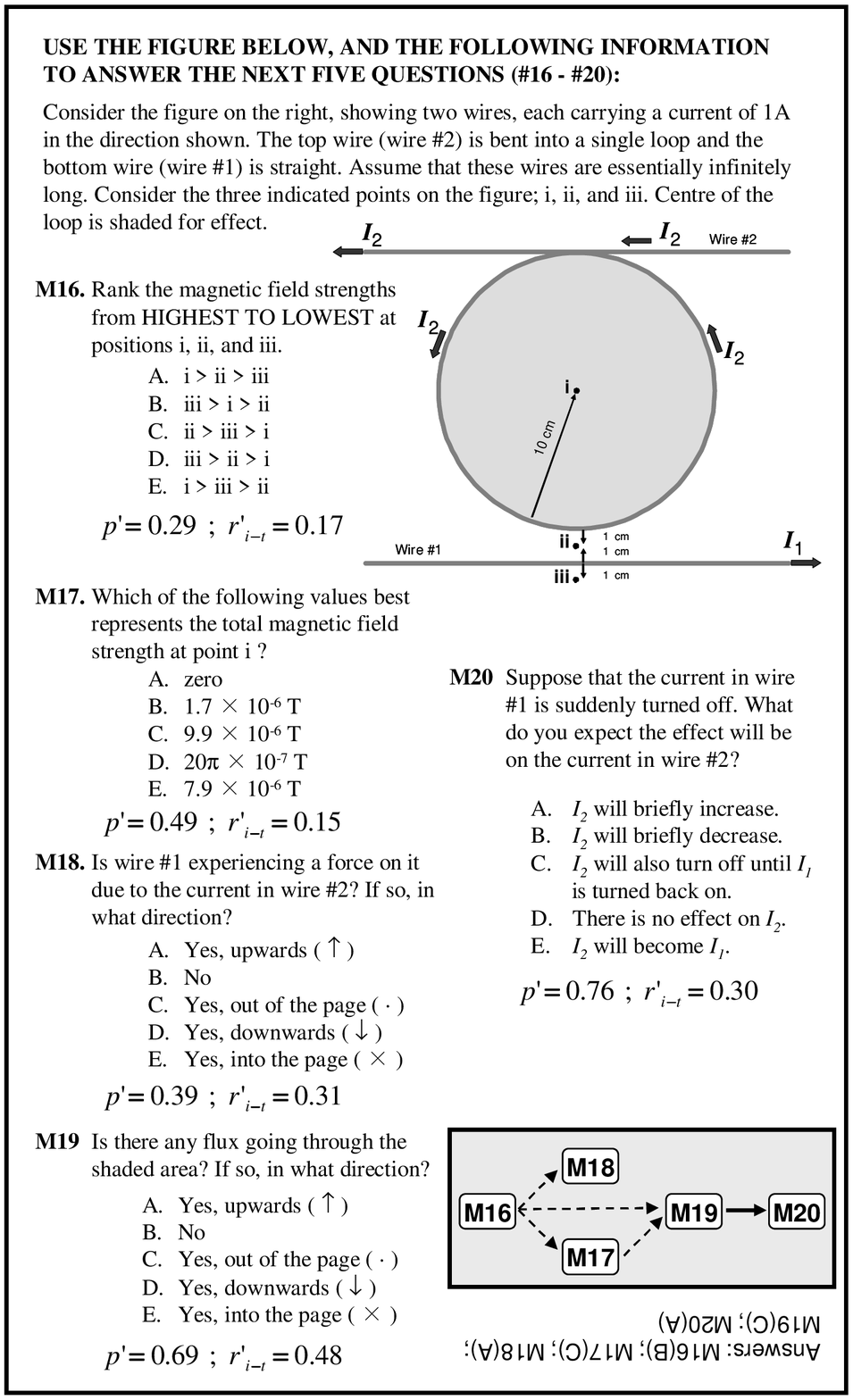}
 \caption{Five-item testlet comprising midterm examination questions M16-M20, representing a moderately integrated item set in which the order of questions is important. Note that the answers from M16 and M17 weakly inform the answer to M19, which in turn is \emph{needed} in order to answer M20.}\label{fig:M16-M20_NewSize2}
\end{figure}

One of the main goals of using MC integrated testlets as stand-ins for CR questions is the detangling of various concepts and tasks into several compartmentalized items that can adequately assess one or two concepts in isolation of the others. With the ability to integrate items within a testlet, one can then re-assemble these items in a given order. As an example, consider the 5-item testlet presented in Figure~\ref{fig:M16-M20_NewSize2},
comprising midterm questions M16-M20. Here, as outlined in the integration map within the figure, there is a complex relationship between the questions. Some questions, such as M16, M17 and M18 are only weakly integrated with each other. The order in which they are presented should not strongly affect how the questions are solved, but the immediate feedback obtained in solving M16, for example, is expected to be of some help to the solving of M17, M18, and M19. On the other hand, the solution to M20 is strongly dependent on the solution to M19. This does not mean that M20 could not be asked without first asking M19. Rather, M19 represents an interpolation of ideas required for M20, and knowledge of this particular concept can be assessed independently of M20. Specifically, M20 tests direct understanding of Lenz's law of induction, while M19 independently tests the concept of magnetic flux that is integral to an understanding of the meaning of Lenz's law. Another way to look at this testlet is to consider questions M16-M18 as both testing and providing the conceptual ``scaffolding" \cite{Ding_Scaffolding_2009} students often need in order to demonstrate the more integrated synthesis required by traditional CR questions. Furthermore, questions M16 and M17 can be viewed as an example of the qualitative-quantitative isomorphic problem pairs motivated by Singh,\cite{Singh_IPP_2008} but here implemented with the requisite inter-item feedback.

\begin{figure}
 \includegraphics[width=\columnwidth]{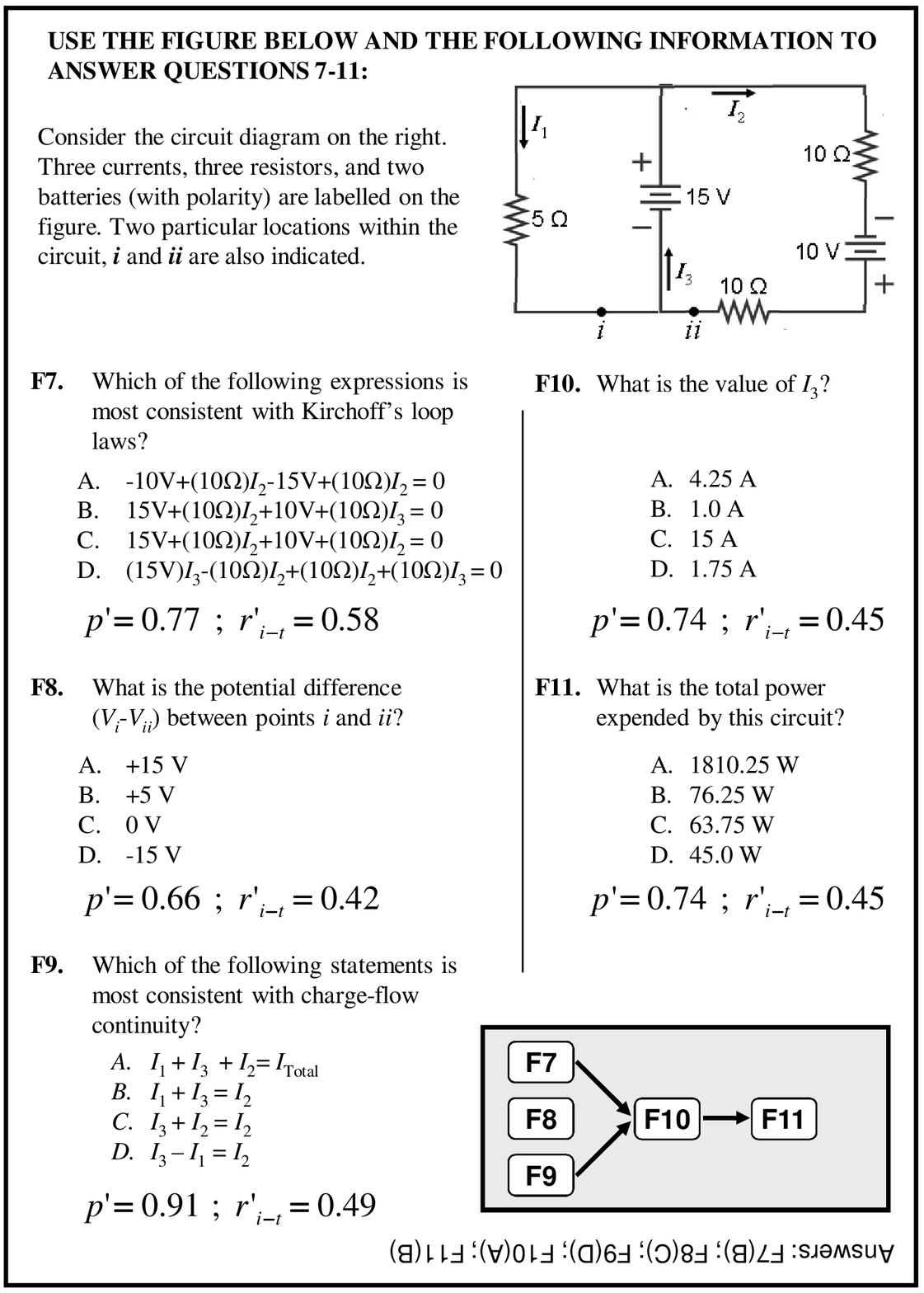}
 \caption{Five-item testlet comprising final examination questions F7-F11, representing a strongly integrated item set in which the order of questions is important. The fact that question F8, which is independent of the other items in the testlet, is the most difficult question in the set may come as a surprise to many readers. All five items in the testlet are strongly discriminating.}\label{fig:F7-F11_NewSize}
\end{figure}

As a final example of a strongly-integrated testlet, consider Figure~\ref{fig:F7-F11_NewSize},
comprising final exam questions F7-F11. This testlet mirrors a common type of CR question involving the application of Kirchoff's circuit laws to simple circuit analysis. The solution of such a question often involves the application of a standard progression of concepts such as Kirchoff's loop law (for voltage), Kirchoff's junction law (for currents), and then an algebraic resolution of the resulting set of equations. In such a CR question, many students will often make an error in the first step and will thus be unable to obtain the proper desired numerical results. This leads to considerable difficulties in grading, as the instructor has to assess the veracity of the students' conceptual and procedural knowledge in a solution that is full of wrong answers. However, with an integrated testlet, these concepts and procedures can be assessed individually. For example, F7 assesses understanding of the loop law, F10 assesses understanding of the junction law, and F9 assesses the algebraic resolution of these laws, while leaving it up to the student to fill in other needed expressions. Once the student has knowledge of one of the currents in the circuit (via the immediate feedback of F10), question F11 can delve deeper into the subject matter and inquire about the power expended by the circuit. Thus, solution to items F7 and F9 are needed for solving F10, which is then needed for solving F11. Item F8 is a stand-alone interpolation meant to assess students' conceptual understanding of voltages as potential \emph{differences}, and is not dependent on the other questions. All five questions prove highly discriminating. Furthermore, as a whole, this 5-item testlet has a fine marking scheme with 21 distinct possible scores ranging nearly uniformly from 0 to 5. Thus, this testlet allows for independent assessment of key concepts and skills in simple circuit analysis and is an excellent proxy for traditional CR questions, but with considerably simpler and more consistent scoring. This integrated testlet is designed to represent the closest proxy to a traditional CR exam question, yet despite its attributes, a key difference between the integrated-testlet and CR approaches remains. The integration of ideas is not only sequential, but also prescriptive. The problem-solving path is contextually (or some might say explicitly) outlined to the student, and thus the testlet provides little opportunity to assess key problem-solving strategies that are inherit to traditional exam problems. The importance of such differences needs to be elucidated with further discussion and research.     

\subsection{Students' perceptions and attitudes}
As outlined above, the IF-AT possesses numerous attendant procedural and pedagogical benefits for classroom testing in introductory physics. Fortunately, this technique has also proven to be extremely attractive to students, who largely recommend its adoption for a wide range of classroom assessments and exams [16]. Students generally find the IF-AT more fun and more fair than other techniques and report feeling better engaged with this technique than with traditional MC formats. When surveyed after the two-hour midterm examination about their experience with the IF-AT technique students showed that in addition to its procedural advantages they also appreciate the pedagogical advantages of the technique. The attitudes towards the IF-AT of the responders were categorized as ``highly-positive'', ``positive'', ``neutral'', ``negative'', and ``highly-negative''. Of the 26 responders in this anonymous course evaluation survey, thirteen highly-positive, ten positive, two neutral, one negative, and no highly-negative impressions were recorded. Many students identified the knowledge-of-results aspect as a primary benefit of the technique, yet several students specifically mentioned the corrective feedback aspect of the test and the manner in which this technique can be used for creating integrated testlets as major advantages. As one student remarked: ``knowing your mark as you leave the test is great; having the feedback during the test is better''.

\subsection{Limitations of the study and future directions}
This pilot study was designed to establish the feasibility of using an answer-until-correct MC response system for creating a new type assessment tool in introductory physics; namely, the ``integrated testlet". In the integrated testlet, a set of related MC items are inter-dependent at various levels, and thus rely on immediate feedback for valid (and fair) progression within an overarching problem. The primary motivation for this lies in the assumption that integrated testlets better approximate traditional constructed-response problems than do traditional non-integrated MC items. Literature on how traditional MC and CR questions relate cognitively is extremely limited in the physics education community.  Studies have found that MC tests of low-level physics knowledge (physics definitions, for example) are sufficiently reliable to replace their CR counterparts, yet they are found to only test limited common cognitive elements \cite{Kruglak_1965}. Thus, while the IF-AT examinations are found reliable and discriminating in this study, the direct comparison of integrated-testlets and CR items needs to be conducted in order to demonstrate that the hypothesized relationship between the two manifests in practice. At that time, one outstanding difference between the two approaches---for example, that of the prescriptive nature of integrated-testlets---will have to be addressed. In additional avenues of research, one could show that non-integrated MC testlets operate fundamentally differently than do integrated testlets. Such studies are crucial for solidifying the findings of this pilot study, and will require careful design and implementation, as well as a much larger dataset. Another proposed avenue of future research is explicit confirmation of the formative nature of the IF-AT-administered integrated-testlet approach. While there is evidence in the literature that the IF-AT promotes learning in other assessment circumstances and disciplines,\cite{DiBattista2005} future experiments will need to be designed to specifically demonstrate that learning is, in fact, taking place during our tests. There is some fear (by at least one reviewer) that the answer-until-correct approach may simply be inflating exam scores exclusive of any perceived formative benefits. The fact that partial credit is being reliably granted to the better students is circumstantial, but certainly not direct, evidence of during-exam learning. Thus, future experiments will have to be carfully designed to eluciate the relationship between formative and summative roles of answer-until-correct integrated-testlet-based exams. Finally, while all of the statistics reported in this article have sufficient statistical power, this study only reports on the results from one (N=60) introductory physics course. We have, in fact, successfully deployed IF-AT-based formal examinations utilizing integrated testlets in three separate introductory physics courses (N=250 in total), but for clarity chose to describe the results from one self-contained case-study. It remains to be shown formally that the approach presented here is applicable to other courses and other institutions.

\section{Summary}
Multiple-choice testing is making up an ever-increasing portion of in-class assessments in introductory physics education. Standard MC formats are sometimes seen as deficient in their ability to assess broadly across the required cognitive taxonomy of introductory physics and deficient in their ability to assess partial knowledge of the material. Answer-until-correct test formats, such as the commercially-available immediate-feedback assessment technique, possess many benefits that allow them to be used procedurally much like a standard MC technique, but pedagogically much more like a constructed-response technique. With the IF-AT, students get corrective or confirmatory feedback \emph{during} the test, and thus, not only leave with full knowledge of their score, but they also learn much about the solution of the test items while they still have an opportunity to incorporate and utilize this knowledge in subsequent questions. This feature further allows for the development of \emph{integrated testlets} in which questions build one upon another, much like concepts in a constructed-response questions. The answer-until-correct format also allows for seamless integration of partial credit that assesses proximal knowledge, with the benefit that a students' final answer is always the fully correct one. Two formal MC examinations comprising both stand-alone questions and testlets of various levels of integration were administered via the IF-AT in an introductory physics university class. These forty-five MC items displayed a high level of item discrimination and provided good overall test reliabilities. The straightforward allocation of partial credit was shown to be valid and discriminating. The ability to incorporate partial-credit in a MC test format is both beneficial pedagogically and increases the attractiveness of the technique to students. The IF-AT was designed as a valuable tool for standardized and classroom testing in the social sciences, but because it enables the practical integration of testlet items, it becomes uniquely attractive in physics education where conceptual scaffolding and integration are keys to developing and assessing physics synthesis and analysis skills.

\begin{acknowledgments}
I thank David DiBattista of Brock University for invaluable insight, guidance, and extensive manuscript editing. I thank Ralph Shiell for critical analysis, occasional antagonism, and strong support of this project. I thank Angie Best of the Instructional Development Centre for financial support. I thank James Day of the University of British Columbia for expert advice, insight and editing. I would also like to thank students and teaching assistants Mike Becker, Amany Raslan, Michael Tessier, Steven Braithwaite, Brad Simpson, and Bruce Darling for test-driving the IF-AT midterms and finals and for their insightful suggestions. Lisa Ugray assisted with type-setting and LaTeX. I would also like to thank Joss Ives (University of the Fraser Valley), and an anonymous referee for considered and thorough comments, discussion, and manuscript editing.  
\end{acknowledgments}

\bibliography{Slepkov_IF-AT}{}

\begin{thebibliography}{}
\bibitem{MultipleKeys}
There is, however, a variety of IF-AT forms with variety of answer keys
  available from Epstein Educational---a fact that greatly aids in test
  security.
\bibitem{WithoutTopTwo}
The top two scorers overall were excluded from this analysis because they
  scored 100\% on the exam and thus had no opportunity for earning partial
  credit. The next fifteen top-ranked students were used to represent the top
  group.
\bibitem{QuestionOrder}
For non-integrated items it is presumed that the order of question presentation
  is irrelevant. However, there is some indication that this may not be the
  whole story. See: \cite{Gray}.
\bibitem{DiffMatSameMeth}
This ``Introductory Physics for the Life Sciences'' is an algebra-based course
  that significantly differs in subject matter from the ``Introductory Physics
  II: Electricity and Magnetism'' course discussed throughout this article.
  While the topics between the two courses were different, the methods of
  instruction and assessment were nearly identical.
\end{thebibliography}
\bibliographystyle{ajp}

\end{document}